\documentclass{PoS}
\title{Theoretical and practical progresses in the HAL QCD method}

\ShortTitle{Theoretical and practical progresses in the HAL QCD method}

\author{\speaker{Sinya~Aoki}\thanks{For HAL QCD Collaboration}
\\
        Center for Gravitational Physics, Yukawa Institute for Theoretical Physics, Kyoto University,
        Kitashirakawa Oiwakecho, Sakyo-ku, Kyoto 606-8502, Japan\\
        E-mail: \email{asoki@yukawa.kyoto-u.ac.jp}}

\abstract{In this report, we discuss some theoretical and practical progresses in the HAL QCD potential method. 
We first clarify the issue of the derivative expansion for the non-local potential in the HAL QCD method. As the non-local potential in the original literature is not uniquely defined,  we propose a procedure to define a non-local potential from NBS wave functions in terms of the derivative expansion. We then demonstrate how this definition works by using quantum mechanics with a separable potential. 
Secondly we discuss an issue of Hermiticity of the HAL QCD potential.
Since the NBS wav functions are not orthogonal to each other in general,
the HAL QCD potential is necessary to be non-Hermitian. 
We consider the next-to-leading order potential, which can be made 
Hermitian exactly by the change of variables.
In general we can also make the higher order HAL QCD potential Hermitian 
order by order in the derivative expansion. 
An explicit example on how the procedure works is given for lattice QCD calculations.
Finally we discuss how we can extract the HAL QCD potential from the NBS wave function in the boosted system. An explicit formula for this is derived. 
          }
          
          \FullConference{37th International Symposium on Lattice Field Theory - Lattice2019\\
		16-22 June 2019\\
		Wuhan, China}

\begin{document}

\section{Introduction}
The hadron interactions have been investigated mainly by two methods in lattice QCD,
the finite volume method\cite{Luscher:1990ux} and 
the HAL QCD potential method \cite{ Ishii:2006ec,Aoki:2009ji,Aoki:2012tk}.
While the two methods are in principle theoretically equivalent,
results of the scattering phase shifts for two baryons sometimes differ between two methods.
Recently origins of the differences have been clarified in Refs.~\cite{Iritani:2016jie,Iritani:2017rlk}.   

The HAL QCD method employs the Nambu-Bethe-Salpeter (NBS) wave function, define as, for example for $NN$, 
\begin{eqnarray}
\varphi^{\bf k}({\bf x}) e^{-W_{\bf k} t} &=& \langle 0\vert N({\bf r},t) N({\bf r} +{\bf x},t) \vert NN, W_{\bf k}\rangle, \quad W_{\bf k} = 2\sqrt{{\bf k}^2 + m_N^2}, 
\end{eqnarray}
whose asymptotic behavior at large $x=\vert{\bf x}\vert$ is given by\cite{Lin:2001ek, Aoki:2005uf, Ishizuka:2009bx, Aoki:2009ji}
\begin{eqnarray}
\simeq \sum_{lm} C_{lm} \frac{\sin(kx +l\pi/2 +\delta_l(k))}{kx} Y_{lm}(\Omega_{\bf x}),
\end{eqnarray}
where $\delta_l(k)$ is the scattering phase shift for the $l$-th partial wave as a function of $k=\vert{\bf k}\vert$.
An energy-independent and non-local potential is defined from the NBS wave functions to satisfy
\begin{eqnarray}
(E_{\bf k} - H_0) \varphi^{\bf k}({\bf x}) &=& \int d^3y\, U({\bf x}, {\bf y}) \varphi^{\bf k}({\bf y}), \quad
E_{\bf k} =\frac{{\bf k}^2}{m_N}, \quad H_0 =-\frac{\nabla^2}{m_N},
\label{eq:def_pot}
\end{eqnarray}
for $^\forall W_{\bf k} \le W_{\rm th}$, where $W_{\rm th} \equiv 2m_N + m_\pi$ is the inelastic threshold. In practice, the non-locality is dealt with the derivative expansion as
$U({\bf x}, {\bf y}) = V({\bf x},\nabla) \delta^{(3)}({\bf x}-{\bf y})$, where
\begin{eqnarray}
V({\bf x},\nabla) &=& V_0(x) + V_\sigma(x) (\sigma_1\cdot \sigma_2) + V_T(x) S_{12} 
+V_{\rm LS}(x) {\bf L}\cdot {\bf S} +O(\nabla^2) .
\end{eqnarray}

We discuss some progresses on three of theoretical and practical issues for the HAL QCD potential method, which are related to the validity of the derivative expansion, non-Hermiticity of the HAL QCD potential and the HAL QCD potential in the moving systems.

\section{Definition of the HAL QCD potential with the derivative expansion}
Eq.~(\ref{eq:def_pot}) does not fix the non-local potential uniquely, due to the restriction of energy $W_{\bf k} \le W_{\rm th}$. Therefore, we have to give the definition (or scheme) of the potential without ambiguity. We here propose one scheme to define a potential completely using the derivative expansion.

For a simplicity of explanations, we consider a potentials between two scalar particles.  
We expand the rotational symmetric potential in terms of $\nabla^2$ and ${\bf L}^2$.
Another rotational symmetric term $({\bf r}\cdot {\bf L})^2$ can be expressed by using $\nabla^2$ and ${\bf L}^2$. Since the NBS wave functions are not orthogonal to each other, the potential should be non-hermitian in general. Thus terms with odd number of $\nabla$ are possible.
However, we do not include such terms in our scheme of the potential. 
Of course one may use other schemes with such terms instead.
 
In our scheme, the rotationally invariant potential is expanded as
\begin{eqnarray}
V({\bf x},\nabla) &=& \sum_{n=0}^\infty\sum_{j=0}^n V_{j, n-j}(x) (\nabla^2)^j ({\bf L}^2)^{n-j},
\end{eqnarray}
where $\nabla^2$ and ${\bf L}^2$ commute with each others, so that their orders are irrelevant,
while we alway put $V_{j,n-j }(x)$ in the left of all $\nabla^2$'s.
Since $V_{j,n-j }(x)$ does not commute with  $\nabla^2$, terms with $j > 0$ are in general non-Hermitian.
In order to determine $V_{j,n-j }(x)$, we first consider $N$-th approximation, defined by
\begin{eqnarray}
V^{(N)}({\bf x},\nabla) &=& \sum_{n=0}^N \sum_{j=0}^n V_{j, n-j}^{(N)}(x) (\nabla^2)^j ({\bf L}^2)^{n-j},
\end{eqnarray}
and determine  unknown functions $V_{j, n-j}^{(N)}(x)$ from 
\begin{eqnarray}
\sum_{n=0}^N \sum_{j=0}^n V_{j, n-j}^{(N)}(x) (\nabla^2)^j ({\bf L}^2)^{n-j} \phi^{{\bf k}_i}({\bf x}) =
\left(\frac{{\bf k}_i^2}{m} -H_0\right)  \phi^{{\bf k}_i}({\bf x})
\end{eqnarray}
for $i=0,1,2,\cdots, N_{\rm total}\equiv N(N+3)/2$, where
$\phi^{{\bf k}_i}({\bf x})$ is the NBS wave function, $m$ is a mass of the scalar particle, 
and ${\bf k}_i$ satisfies $2\sqrt{{\bf k}_i^2 + m^2}=2m+i (W_{\rm th} - 2m)/N_{\rm total}$
for  the inelastic threshold  $W_{\rm th}$. 
Since a number of unknown functions is equal to  a number of equations, $N_{\rm total} +1$, 
$V_{j, n-j}^{(N)}(x)$ can be determined in principle.\footnote{In practice, higher oder terms are numerically difficult to determine.}
Then, we define $V_{j, n-j}(x)= \lim_{N\rightarrow\infty}V_{j, n-j}^{(N)}(x)$.

In order to see how the above procedure works, we consider the quantum mechanics with a separable potential given by
\begin{eqnarray}
U({\bf x}, {\bf y}) &=& \omega v({\bf x}) v({\bf y}), \qquad 
v({\bf x}) = e^{-\mu x},
\end{eqnarray}
which is highly non-local. We solve the Schr\"odinger equation with this potential by introducing an infra-red cut-off $R$. The S-wave ($L=0$) wave function at $k=\vert{\bf k}\vert$ is thus given by
\begin{eqnarray}
\psi_k(x) &=& \left\{
\begin{array}{ll}
\displaystyle
 \frac{e^{i\delta(k)}}{k x} \left[ \sin(kx+\delta(k)) - \sin\delta(k) e^{-\mu x} \left( 1+ x\frac{\mu^2+k^2}{2\mu}\right)\right], %\quad 
&x \le R, \\
\\
\displaystyle
 C(k)  \frac{e^{i\delta(k)}}{k x} \sin (kx+\delta_R(k) ),  %\quad 
& x > R, \\
\end{array}
\right.
\end{eqnarray}
where
\begin{eqnarray}
k\cot \delta(k) &=&-\frac{1}{4\mu^2}\left[2\mu(\mu^2-k^2) -\frac{3\mu^2+k^2}{4\mu^3}(\mu^2+k^2)^2
+\frac{(\mu^2+k^2)^4}{8\pi m\omega}\right],  \\
k\cot \delta_R(k) &=& k \frac{X+\cot(kR) Y}{X \cot(kR) -Y},
\quad X=\psi_k(R),\ Y=\left. \frac{d}{d x} \left(x\psi_k(x) \right)\right\vert_{x=R} ,
\end{eqnarray}
and $m$ is a mass in the Schr\"odinger equation (but not the reduced mass).

In our procedure, we determine the leading order (LO) potential from one wave function as
%\begin{eqnarray}
$V^{(0)}_{0,0}(x) = (E_k - H_0) \psi_k(x)/\psi_k(x)$ with $E_k =k^2/(2m)$.
%\end{eqnarray}
We consider two cases, $k^2=0$ (the lowest threshold) and $k^2=\mu^2$,  which corresponds to
the approximated inelastic threshold in field theory, where $\mu$ may be regarded as a mass of exchanging particles.   By combining two wave functions at $k^2=0$ and $k^2=\mu^2$, we can also
determine the next-to-leading order (NLO) terms for the S-wave, $V^{(1)}_{0,0}(x)$ and $V^{(1)}_{1,0}(x)$.
We calculate the scattering phase shifts using three potentials, two at LO and one at NLO, and compare them with exact one, $\delta_R(k)$.

\begin{figure}[bth]
\centering
%\vskip -2cm
  \includegraphics[angle=0, width=0.49\textwidth]{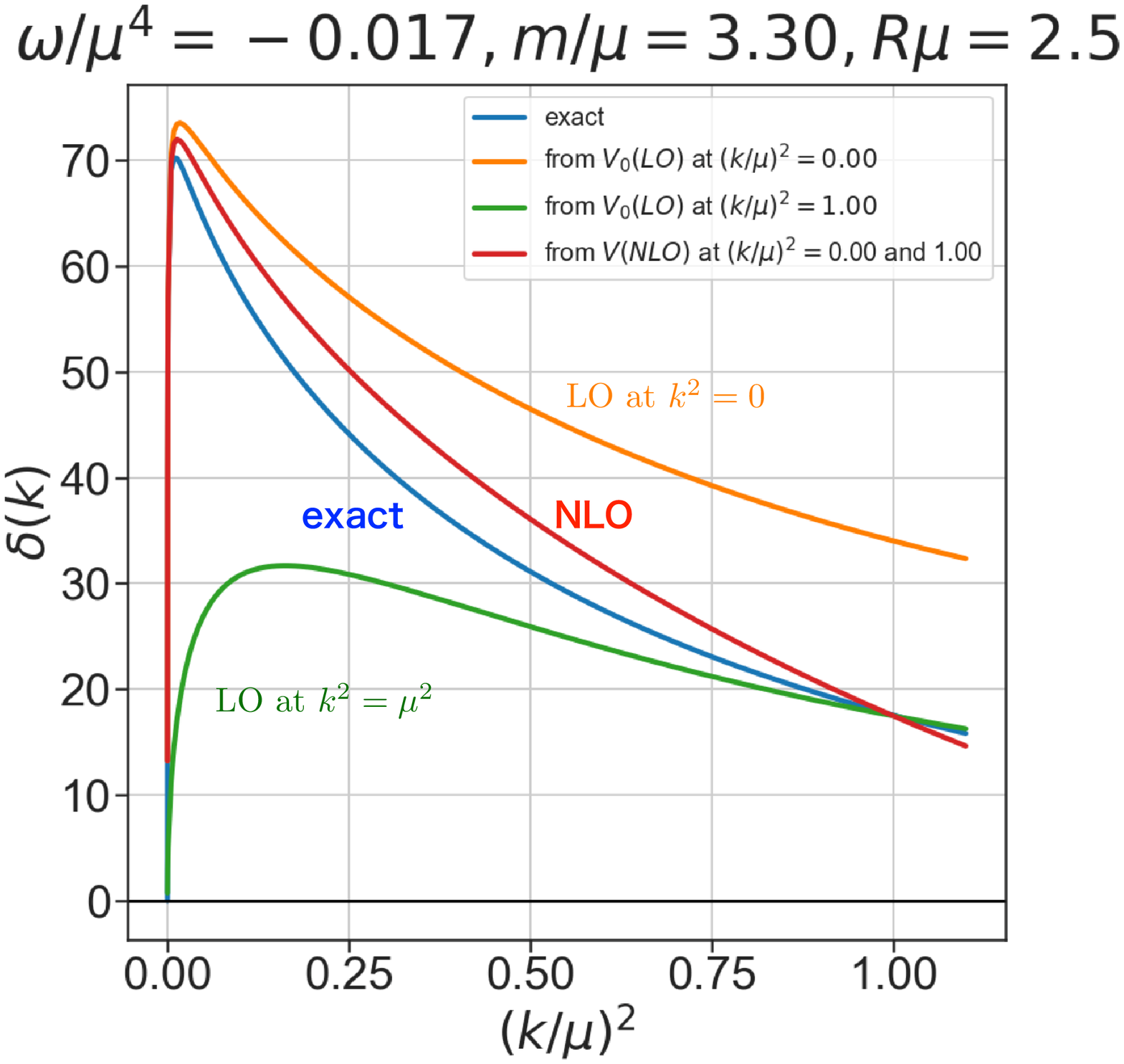}
  \includegraphics[angle=0, width=0.49\textwidth]{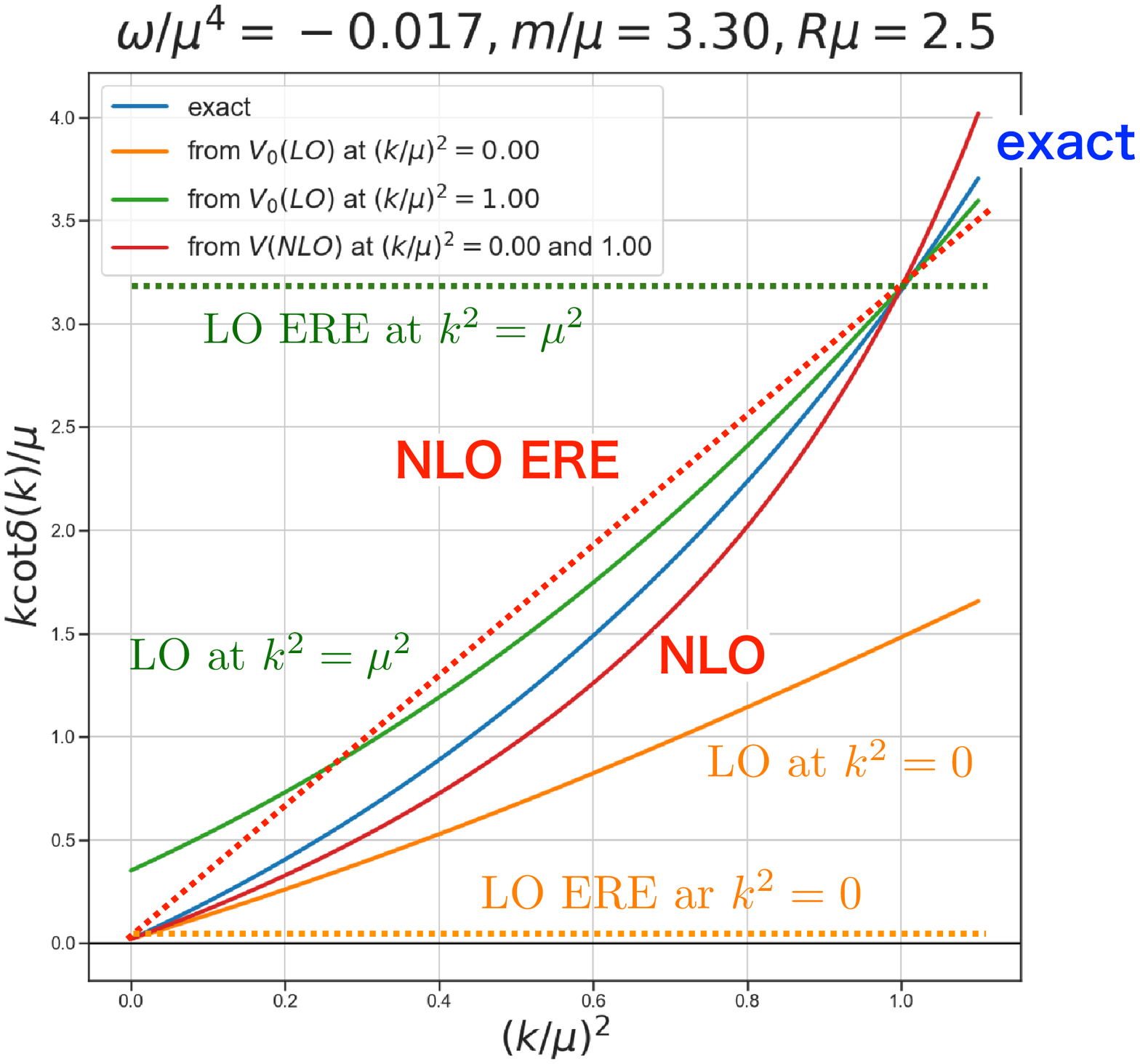} 
 \caption{ Scattering phase shifts $\delta_R(k)$ (Left) and $k\cot\delta_R(k)/\mu$ (Right)
 as a function of $(k/\mu)^2$.
 Exact (blue),  LO at $k^2=0$ (orange), LO at $k^2=\mu^2$ (green) and
 NLO (red).}
 \label{fig:ERE}
\end{figure}
Fig.~\ref{fig:ERE} (Left) compares scattering phase shifts among
the exact (blue), the LO at $k^2=0$ (orange). the LO at $k^2=\mu^2$ (green) and the NLO (red),
at $\omega/\mu^4 = -0.017 $, $m/\mu = 3.3 $ and $R \mu = 2.5$,
which are so chosen that the $k$ dependence of the exact scattering phase shift $\delta_R(k)$ (blue) is similar to the one for the scattering in the $NN$ ($^1S_0$) channel with $m\simeq m_N/2$ and $\mu \simeq m_\pi$.
By construction, the phase shift from the LO potential obtained at $k^2=0$ (orange) agrees with the exact $\delta_R(k)$ (blue) at $k^2=0$ and gradually deviates as the $k^2$ increases toward $k^2=\mu^2$, while 
the LO scattering phase shift from the one at $k^2=\mu^2$ (green) reproduces the exact one (blue) at $k^2=\mu^2$ but gradually deviates as the energy decreases except $k^2=0$ where all scattering phase shifts should become zero. 
On the other hand, the NLO scattering phase shift (red) agrees with  exact values at both $k^2=0$ and $k^2=\mu^2$, and give a reasonable interpolation of the exact $\delta_R(k)$ between $k^2=0$ and $k^2=\mu^2$, as seen from the figure.
These features can be seen more clearly in Fig.~\ref{fig:ERE} (Right), where
$k\cot\delta_R(k)/\mu$ is plotted as a function of $(k/\mu)^2$ for all 4 cases.
If we increase the order of the expansion more and more, the approximated scattering phase shift become closer and closer to the exact one.

Now let us compare results from the HAL QCD method with those from the direct method.
At small $k^2$, the effective range expansion (ERE) reads  
\begin{eqnarray}
k \cot\delta_R(k) &=& \frac{1}{a_0} + \frac{r_{\rm eff}}{2} k^2 + O(k^4),
\end{eqnarray}
where $a_0$ is the scattering length and $r_{\rm eff}$ is the effective range.
If one obtain the scattering phase shift at only one value of $k^2$ in the direct method, one can determine $a_0$, the LO term in the effective range expansion.
In Fig.~\ref{fig:ERE} (Right), the orange and green dotted lines correspond to the LO ERE lines
determined at $k^2=0$ and  $k^2=\mu^2$, respectively.  
Combining two, $a_0$ and $r_{\rm eff}$ can be determined, so that the NLO ERE line is given by the red dotted line. We notice in Fig.~\ref{fig:ERE} (Right) that
$k\cot\delta_R(k)$ obtained with the NLO potential (red solid line) gives a better approximation
of the exact one (blue) than the NLO ERE line (red dotted line).

\section{Construction of Hermitian potential}
Since the NBS wave functions are in general not orthogonal to each other, the HAL QCD potential defined in terms of the derivative expansion is necessary to be non-Hermitian except the LO term (local potential). However we can make non-Hermitian potential Hermitian, as shown in this section. 

\subsection{Formulation}
We consider the Schr\"odiner equation,
\begin{eqnarray}
H \psi &=& E\psi, \quad H= H_0 +U, \quad H_0 \equiv -\frac{1}{2m} \nabla^2,
\end{eqnarray}
where $U$ is non-Hermitian but eigenvalue $E$ is real.
By a change of wave function as $\psi = R \phi$, we have
\begin{eqnarray}
\tilde H \phi &=& E\phi, \quad \tilde H\equiv R^{-1} H R = H_0 + V.
\end{eqnarray}
We thus determine $R$ so as to make $V$ Hermitian.

Let us take the lowest non-trivial example, whose potential is given by
\begin{eqnarray}
U(\vec x) &=& V_0(x) + V_2 (x) \nabla^2 = V_0(x) + \nabla^i V_2(x) \nabla_i
+\delta V(x) \hat x^i \nabla_i,
\end{eqnarray}
where $\hat x^i =x^i/x$ and  $\delta V(x) = - V_2^\prime (x)$, and the last term in the second equation is a non-hermitian part. Then we obtain
\begin{eqnarray}
\tilde H &=& H_0 +\tilde V_0 + \nabla^i V_2 \nabla_i
+\left\{\delta V-  \frac{1-V_2}{m} \frac{d \log R}{d x}\right\}.
\end{eqnarray}
A condition for the last term to vanish can be easily solved as
\begin{eqnarray}
R(x) &=& \exp\left[ m\int_{x_\infty}^x d s\frac{\delta V(s)}{1- 2m V_2(s)}\right], 
\end{eqnarray}
where $x_{\infty}$ represents some distance beyond which $V_2$ and $\delta V$ vanish,
and the LO term is given by
\begin{eqnarray}
\tilde V_0(x) &=& V_0(x) -\frac{\delta V(x)}{x} -\frac{d}{dx} \left(\frac{\delta V(x)}{2}\right)
+\frac{m}{2}\frac{\delta V(x)^2}{(1-2m V_2(x))} .
\end{eqnarray}

Higher order terms such as $V_{2n} (x) (\nabla^2)^n$ can be made Hermitian approximately within the derivative expansion, where $O((\nabla^2)^{n+1})$  contributions are neglected as ``higher order" contributions.
See Ref.~\cite{Aoki:2019gqt} for more details.

\subsection{Example}
 In Ref.~\cite{Iritani:2018zbt}, the NLO potential has been extracted for  $\Xi\Xi (^1S_0)$ system 
 in 2+1 flavor lattice QCD at $a=0.09$ fm and $m_\pi = 510$ MeV.
 \begin{figure}[tbh]
\centering
%\vskip -2cm
  \includegraphics[angle=270, width=0.49\textwidth]{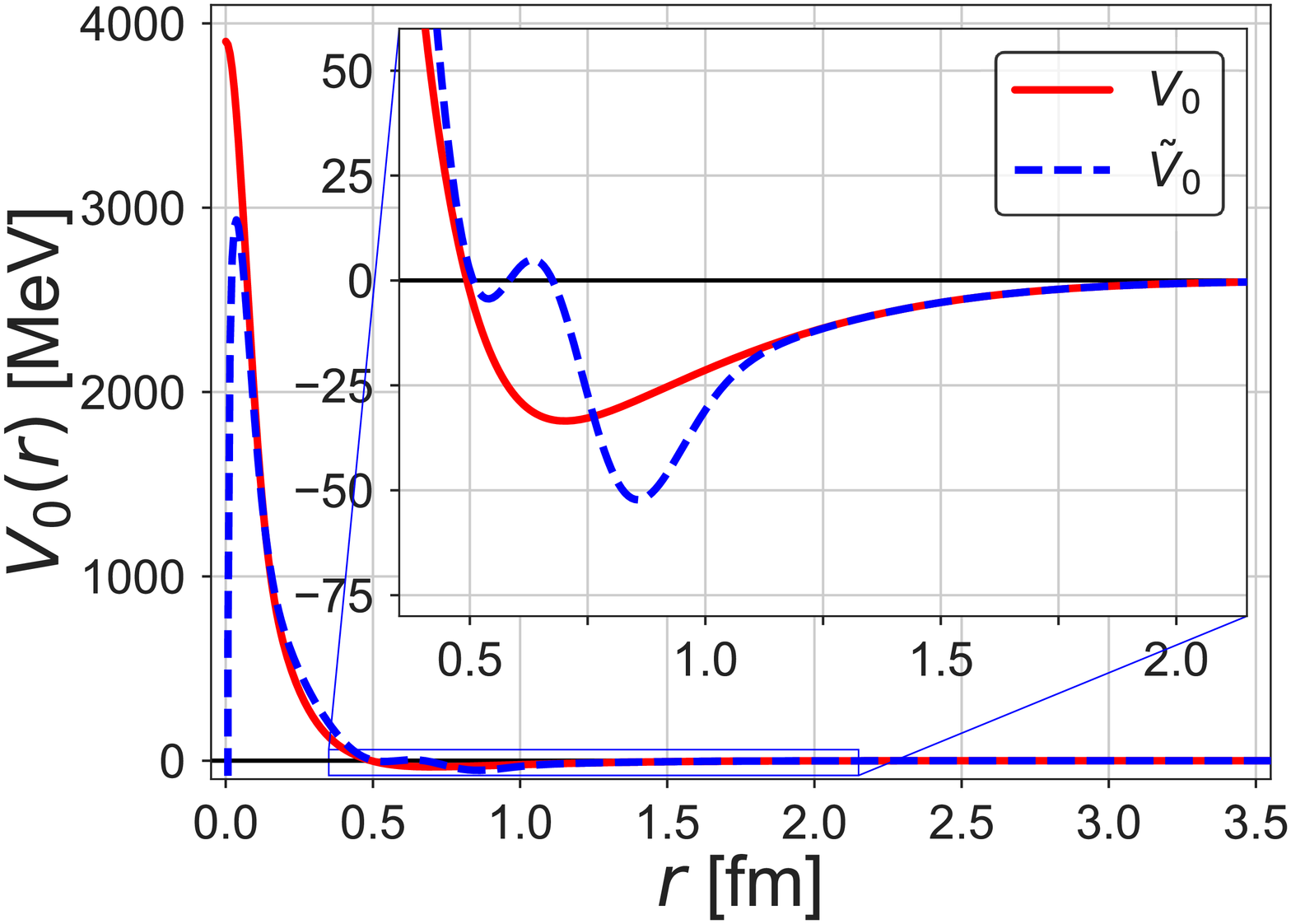}
  \includegraphics[angle=270, width=0.49\textwidth]{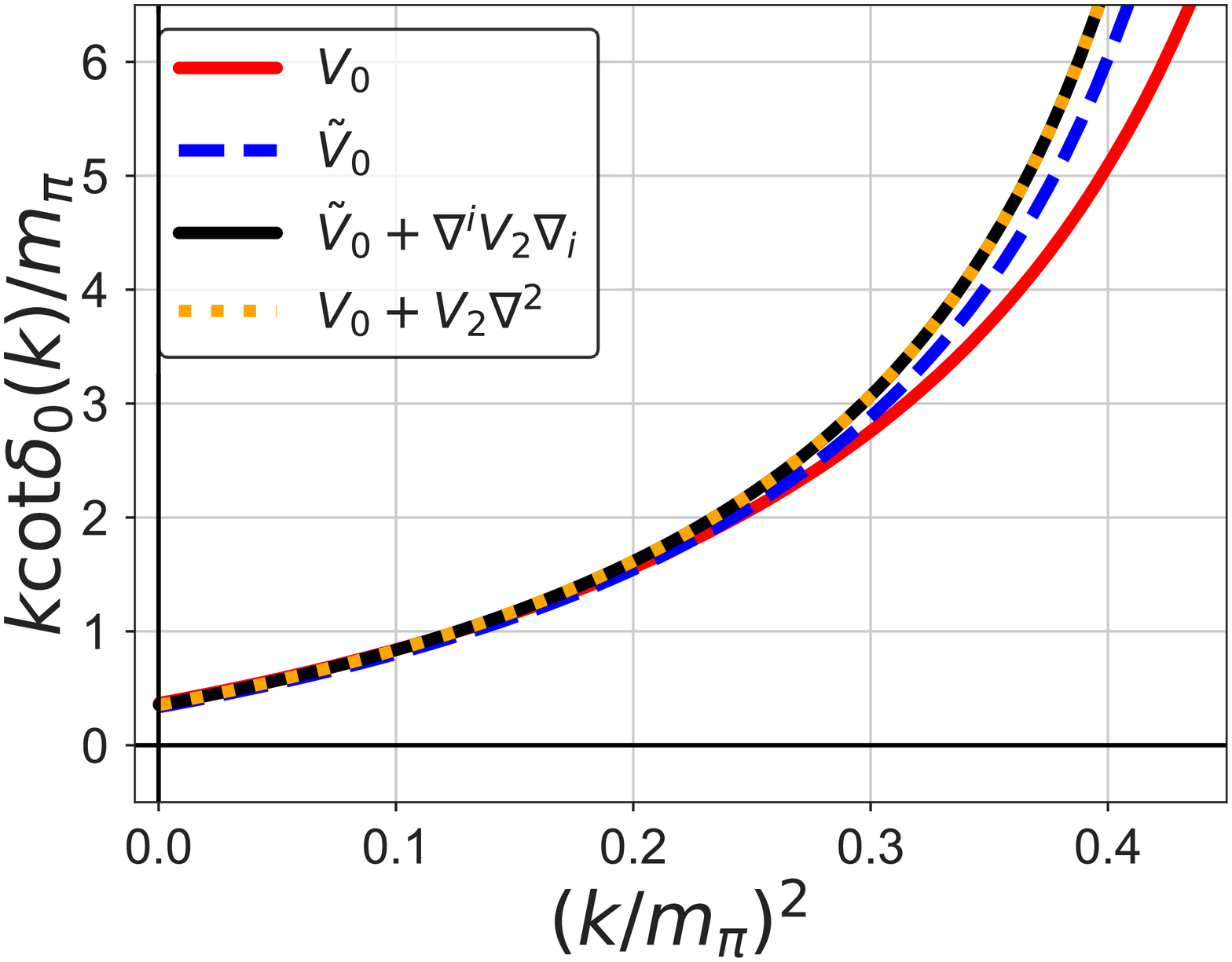} 
 \caption{ (Left) The LO term $\tilde V_0$ (blue dashed line) after Hermitization
 in the NLO potential for $\Xi\Xi (^1S_0)$, together with the original LO term $V_0$ (red solid line). 
  (Right) $k \cot\delta_0(k)/m_\pi$ as a function of $(k/m_\pi)^2$, where
 $\delta_0(k)$ is the scattering phase shift,
 calculated from the NLO potential $V_0 + V_2\nabla^2$ (red solid line) and 
 $\tilde V_0 + \nabla^i V_2\nabla_i $ (yellow dotted line), and those  from the LO term
 $V_0$ (red solid line) and $\tilde V_0$ (blue dotted line). }
 \label{fig:NLO}
\end{figure}

 Using the formula in the previous subsection, we then construct $\tilde V_0$, which is plotted in
 Fig.~\ref{fig:NLO} (Left), together with the original LO term $V_0$ in the NLO potential $V_0 + V_2\nabla^2$. We observe that $\tilde V_0$ shows a more complicated behavior than $V_0$.
 In Fig.~\ref{fig:NLO} (Right), we show $k\cot\delta_0(k)/k$ as a function of $(k/m_\pi)^2$,
 where $\delta_0(k)$ is the scattering phase shift for $\Xi\Xi (^1S_0)$ system calculated from these potentials.
 As seen from the figure, $V_0 +  V_2\nabla^2$ (red solid line) and 
 $\tilde V_0 + \nabla^i V_2\nabla_i $ (yellow dotted line) give identical results
 by construction, showing that the method in the previous subsection indeed works.
The LO terms $V_0$ and $\tilde V_0$ alone also show reasonably good results at low energy but effects of the NLO contributions gradually show up as energy increases.
 Interestingly, the LO term $\tilde V_0$ after Hermitization gives better approximation than 
 $V_0$ before Hermitization at higher energies.
 We however do not have a good explanation for this difference.
 
\section{The HAL QCD potential from the moving system}
To investigate the $\sigma$ resonance precisely in lattice QCD,
we had better to reduce a contamination of the vacuum state, which has same quantum numbers of the $\sigma$  in the center of mass system.
A standard solution is to employ the moving systems, which do not couple to the vacuum state.
However, the HAL QCD method is so far defined in the center of mass system.
Therefore,  in this section, we consider how the moving system can be utilized for the HAL QCD method.

For this purpose, we introduce a generalized NBS wave function for scalar fields in the center of mass system, 
\begin{eqnarray}
\varphi^{{\bf k}^*}({\bf x}^*, x_4^*) &=& \langle 0\vert N({\bf x}^*/2,x_4^*/2) N(-{\bf x}^*/2,-x_4^*/2) \vert NN, W_{{\bf k}^*}\rangle, \quad W_{{\bf k}^*} = 2\sqrt{({\bf k}^*)^2 + m_N^2},
\end{eqnarray}
whose asymptotic behavior at $x\rightarrow\infty$
is %shown to be 
similar to the equal-time NBS wave function as 
\begin{eqnarray}
\varphi^{{\bf k}^*}({\bf x}^*, x_4^*)&\simeq & \sum_{lm} D_{lm} \frac{\sin(k^*x^* +l\pi/2 +\delta_l(k^*))}{k^*x^*} Y_{lm}(\Omega_{{\bf x}^*}) .
\end{eqnarray}
Therefore we can define the HAL QCD potential in the non-zero $x_4^*$ scheme as
\begin{eqnarray}
(E_{{\bf k}^*} - H_0) \varphi^{{\bf k}^*}({\bf x}^*,x_4^*) &=& V_{x_4^*}({\bf x}^*, \nabla) \varphi^{{\bf k}^*}({\bf x}^*,x_4^*),
\end{eqnarray}
where $V_{x_4^*=0}$ corresponds to the HAL QCD potential in the equal time scheme.

The Lorentz transformation, which relates the Euclidean coordinate $x$ in the boosted system with velocity ${\bf V}$ to the one $x^*$ in the center of mass system, leads to
$\phi^{{\bf k}^*}( x^*) = \phi^{{\bf k}, k_0}( x)$, where
$x^* = ({\bf x}_\perp, \gamma ({\bf x}_\parallel +i{\bf V} x_4), \gamma(x_4 -i{\bf V}\cdot {\bf x} ))$
and $({\bf k}^*, 0) =({\bf k}_\perp, \gamma({\bf k}_\parallel - {\bf V}{k_0}), \gamma(k_0-{\bf V}\cdot {\bf k}))$ 
with the boost factor $\gamma=1/\sqrt{1-{\bf V}^2}$.
Here $x_\perp$( $x_\parallel$) is perpendicular (parallel) to ${\bf V}$. 
Using the relation, we obtain
\begin{eqnarray}
V_{\gamma(x_4-i{\bf V}\cdot{\bf x})}(x_\perp, \gamma ({\bf x}_\parallel +i{\bf V} x_4), \nabla_{x^*})  \phi^{{\bf k}, k_0}( x)
&=& \frac{({\bf k}^*)^2 + \nabla_{x^*}^2}{m_N}  \phi^{{\bf k}, k_0}( x) ,
\end{eqnarray}
where $\nabla_{x^*}^2 =\nabla_{x_\perp}^2 + \gamma^2(\nabla_{x_\parallel} + i{\bf V} \partial_{x_4})^2$. We thus set $x_4$ in order to avoid imaginary spatial coordinates in the potential.
 The final expression becomes
\begin{eqnarray}
V_{-i\gamma{\bf V}\cdot{\bf x}}(x_\perp, \gamma {\bf x}_\parallel , \nabla_{x^*})  \phi^{{\bf k}, k_0}( x)
&=& \frac{({\bf k}^*)^2 + \nabla_{x^*}^2}{m_N}  \phi^{{\bf k}, k_0}( x) ,
\end{eqnarray}
 which is the potential in the $x_4^* = -i\gamma {\bf x}_\parallel$ scheme
( $x_0^* = -\gamma {\bf x}_\parallel$ for to the Minkowski  time) at each ${\bf x}_\parallel$,
 and the potential in the equal-time scheme is extracted at ${\bf x}_\parallel = 0$.  

It is also possible to apply the time-dependent method for the HAL QCD potential\cite{HALQCD:2012aa} to the boosted NBS wave function. We are now testing the method by applying the above formula to the $I=2$ $\pi\pi$ system in the moving frame.

\vskip 0.5cm 

I would like to thank Drs.~Takumi~Iritani and Koichi~Yazaki for collaborations in a part of this work, and members of the HAL QCD collaboration for useful discussions.
This work is supported in part by the Grant-in-Aid of the Japanese Ministry of Education, Sciences and Technology, Sports and Culture (MEXT) for Scientific Research (Nos. JP16H03978, JP18H05236),  
by a priority issue (Elucidation of the fundamental laws and evolution of the universe) to be tackled by using Post ``K" Computer, and by Joint Institute for Computational Fundamental Science (JICFuS).


\begin{thebibliography}{99}
\bibitem{Luscher:1990ux}
  M.~Luscher,
  %``Two particle states on a torus and their relation to the scattering matrix,''
  Nucl.\ Phys.\ B {\bf 354} (1991) 531.
  
\bibitem{Ishii:2006ec}
Noriyoshi Ishii,  Sinya Aoki and Tetsuo Hatsuda,  Phys. Rev. Lett. 99 (2007) 022001.

\bibitem{Aoki:2009ji}
Sinya Aoki, Tetsuo Hatsuda and Noriyoshi Ishii,  Prog. Theor. Phys. 123 (2010) 89.

\bibitem{Aoki:2012tk}
 S.~Aoki {\it et al.} [HAL QCD Collaboration],
%S.~Aoki, T.~Doi, T.~Hatsuda,  Y.~Ikeda, T.~Inoue, Takashi N.~Ishii, K.~Murano, H.~Nemura and K.~Sasaki, 
PTEP 2012 (2012) 01A105.

\bibitem{Iritani:2016jie}
  T.~Iritani {\it et al.} [HAL QCD Collaboration],
  %``Mirage in Temporal Correlation functions for Baryon-Baryon Interactions in Lattice QCD,''
  JHEP {\bf 1610} (2016) 101.
  
 \bibitem{Iritani:2017rlk}
  T.~Iritani {\it et al.} [HAL QCD Collaboration],
  %``Are two nucleons bound in lattice QCD for heavy quark masses? Consistency check with L\UTF{00FC}scher’s finite volume formula,''
  Phys.\ Rev.\ D {\bf 96} (2017) no.3,  034521.

\bibitem{Lin:2001ek}
  C.~J.~D.~Lin, G.~Martinelli, C.~T.~Sachrajda and M.~Testa,
  %``K --> pi pi decays in a finite volume,''
  Nucl.\ Phys.\ B {\bf 619} (2001) 467.

\bibitem{Aoki:2005uf}
  S.~Aoki {\it et al.} [CP-PACS Collaboration],
  %``I=2 pion scattering length from two-pion wave functions,''
  Phys.\ Rev.\ D {\bf 71} (2005) 094504.  

\bibitem{Ishizuka:2009bx}
  N.~Ishizuka,
  %``Derivation of Luscher's finite size formula for N pi and NN system,''
  PoS LAT {\bf 2009} (2009) 119.
  
  \bibitem{Aoki:2019gqt}
  S.~Aoki, T.~Iritani and K.~Yazaki,
  %``Hermitizing the HAL QCD potential in the derivative expansion,''
  arXiv:1909.00656 [hep-lat].
 
 \bibitem{Iritani:2018zbt}
  T.~Iritani {\it et al.} [HAL QCD Collaboration],
  %``Systematics of the HAL QCD Potential at Low Energies in Lattice QCD,''
  Phys.\ Rev.\ D {\bf 99} (2019) no.1,  014514.
%  doi:10.1103/PhysRevD.99.014514
 % [arXiv:1805.02365 [hep-lat]]. 
  
  \bibitem{HALQCD:2012aa}
  N.~Ishii {\it et al.} [HAL QCD Collaboration],
  %``Hadron\UTF{2013}hadron interactions from imaginary-time Nambu\UTF{2013}Bethe\UTF{2013}Salpeter wave function on the lattice,''
  Phys.\ Lett.\ B {\bf 712} (2012) 437.
 % doi:10.1016/j.physletb.2012.04.076
  %[arXiv:1203.3642 [hep-lat]].
  
      
\end{thebibliography}
\end{document}